\documentclass[
 reprint,
 superscriptaddress,
 amsmath,amssymb,
 aps,
 prd,
]{revtex4-2}

\usepackage{graphicx}
\usepackage{dcolumn}
\usepackage{bm}
\usepackage[colorlinks=true, citecolor=Blue, linkcolor=Navy, urlcolor=RoyalBlue]{hyperref}

\usepackage{multirow} 
\newcolumntype{P}[1]{>{\centering\arraybackslash}p{#1}}
\newcolumntype{M}[1]{>{\centering\arraybackslash}m{#1}}

\usepackage{diagbox}
\usepackage[svgnames,table]{xcolor}

\usepackage{caption}
\usepackage{subcaption}
\usepackage{ragged2e}

\usepackage{braket}
\usepackage{diagbox}  
\usepackage{makecell}

\DeclareCaptionJustification{justified}{\justifying}

\renewcommand{\arraystretch}{1.6} 
\newcommand{\LG}{\mathrm{LG}}
\newcommand{\LGmode}[2]{\ensuremath{\LG_{#1,#2}}}

\begin{document}


\title{Optimizing Laguerre-Gaussian Mode Interferometry in Gravitational-Wave Detectors through Modal Degeneracy Breaking}

\author{Liu Tao}
\email{liu.tao@apc.in2p3.fr}
\affiliation{Universit\'e Paris Cit\'e, CNRS, Astroparticule et Cosmologie, F-75013 Paris, France}

\author{Eleonora Capocasa}
\affiliation{Universit\'e Paris Cit\'e, CNRS, Astroparticule et Cosmologie, F-75013 Paris, France}

\author{Matteo Barsuglia}
\affiliation{Universit\'e Paris Cit\'e, CNRS, Astroparticule et Cosmologie, F-75013 Paris, France}

\date{\today}

\begin{abstract}

Higher-order Laguerre-Gaussian modes can reduce the impact of test-mass thermal noise in gravitational-wave interferometers, but their practical implementation is limited by degeneracy-induced coupling losses in realistic high-finesse cavities. In this work, we mitigate these losses by breaking the modal degeneracy through optimization of the mirror curvature. We develop a generalized framework for designing an axisymmetric corrective profile on the test-mass surface that lifts the degeneracy of the sixth-order Laguerre-Gaussian mode subspace while preserving the quality of an injected donut-shaped $\mathrm{LG}_{0,6}$ mode. The profile is expanded in a Fourier-Bessel basis and optimized to maximize the frequency splitting from the nearest unwanted mode while minimizing the loss of the target mode. For a Virgo-like Fabry-Perot arm cavity, the optimized profile separates the injected $\mathrm{LG}_{0,6}$ mode resonance from the nearest same-order resonance, $\mathrm{LG}_{1,4}$, by $2.04\%$ of the free spectral range, reducing the gain of the unwanted modes below unity at the target-mode resonance and thereby suppressing their resonant enhancement. The profile introduces negligible target-mode loss and improves robustness against astigmatism and realistic mirror surface distortions. Monte Carlo simulations with realistic mirror surface imperfections yield an average contrast defect of $368\,\mathrm{ppm}$ and an average reflected optical loss of $979\,\mathrm{ppm}$, corresponding to an equivalent round-trip loss of approximately $3.5\,\mathrm{ppm}$ for the modeled cavity. These results substantially improve upon the central anti-reflective mirror mask approach based on selective loss engineering, bringing the reflected optical loss within a factor of two of that obtained with the fundamental Gaussian mode under the same simulation conditions. The proposed approach offers a promising route toward robust donut-shaped Laguerre-Gaussian modes in future precision interferometers with realistic imperfections.

\end{abstract}

\maketitle


\section{Introduction}

Higher-order Laguerre-Gaussian (LG) modes have been extensively studied as a promising route to reducing test mass thermal noise in gravitational-wave detectors~\cite{PhysRevLett.105.231102, PhysRevD.90.122011, PhysRevD.92.102002, PhysRevD.82.012002, Bond_2011, PhysRevD.84.102001, Sorazu_2013, HOM_thermal}. Compared with the fundamental Gaussian mode, higher-order LG modes provide broader and more spatially distributed intensity profiles on the test masses, thereby averaging displacement fluctuations over a larger mirror area and reducing the impact of coating and substrate thermal noise~\cite{Mours_2006, Vinet2009zz, PhysRevD.82.042003}. This makes them attractive candidates for future interferometric gravitational-wave detectors, where coating thermal noise remains one of the key limitations to sensitivity~\cite{PostO5Report:2022, CEHorizonStudy, ETDesignReportUpdate2020}. 

\begin{figure}[t]
\centering
\includegraphics[width=1\linewidth]{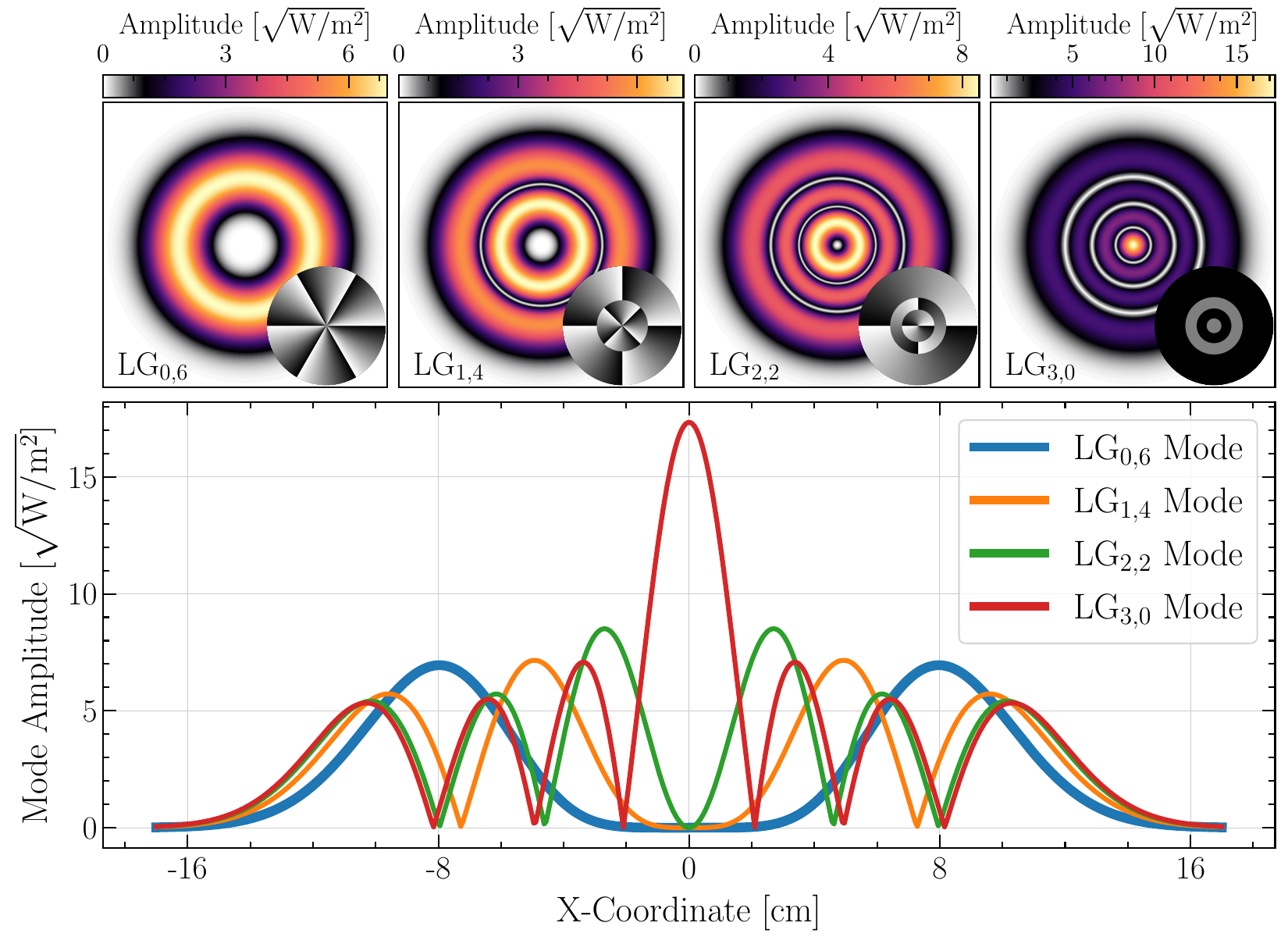}
\caption{Amplitude and phase distributions of the LG modes with mode order $\mathcal{N}=6$. \textit{Top:} Two-dimensional amplitude profiles; the inset in the lower-right corner of each panel shows the corresponding transverse phase profile, highlighting the azimuthal phase winding and radial phase structure. \textit{Bottom:} One-dimensional cross sections of the amplitude profiles along the horizontal axis.}
\label{fig-LG_mode_I_phase}
\end{figure}

More generally, LG modes \LGmode{p}{\ell} form a complete orthonormal basis for paraxial optical fields, with each mode characterized by a radial index $p$ and an azimuthal index $\ell$~\cite{Bond2017}. The corresponding transverse mode order is defined as $\mathcal{N}=2p+|\ell|$. Fig.~\ref{fig-LG_mode_I_phase} shows representative amplitude and phase distributions for \LGmode{p}{\ell} modes with the same order, $\mathcal{N}=6$. Only modes with positive azimuthal index are shown; modes with negative azimuthal index have the same amplitude distribution as their positive-index counterparts, but with the opposite phase winding. For helically phased \LGmode{p}{\ell} beams with $\ell \neq 0$, the transverse phase contains the azimuthal factor $\exp(i\ell\phi)$, where $\phi$ is the azimuthal angle. These modes carry orbital angular momentum, with each photon carrying $\ell\hbar$, where $\ell$ is also known as the topological charge and $\hbar=h/2\pi$ is the reduced Planck constant~\cite{PhysRevA.45.8185}. This additional spatial degree of freedom has made LG modes attractive in a wide range of applications beyond gravitational-wave detection, including optical communications, quantum information, and precision metrology~\cite{Mair_2001, padgett2011tweezers, Wang:2012nkp, Erhard_2017, Fontaine_2019}.

However, a major practical limitation of higher-order LG modes is their susceptibility to imperfection-induced optical losses arising from modal degeneracy~\cite{Bond_2011}. In an ideal spherical optical cavity, all LG modes with the same transverse mode order $\mathcal{N} = 2p + |\ell|$ acquire the same round-trip Gouy phase and are therefore resonant at the same frequency~\cite{Bond2017}. Realistic imperfections, such as mirror surface distortions, can therefore scatter power from the injected higher-order mode into other modes within the same degenerate subspace. Because these scattered modes are also resonant in the cavity, they can be coherently enhanced rather than filtered out, leading to significant degradation of the circulating beam quality and increased optical loss. For example, previous studies have shown that the \LGmode{3}{3} mode is particularly sensitive to residual astigmatism at levels compatible with state-of-the-art mirror surfaces in realistic gravitational-wave detector cavities~\cite{Sorazu_2013, PhysRevD.92.102002}.

\begin{figure}[t]
\centering
\includegraphics[width=1\linewidth]{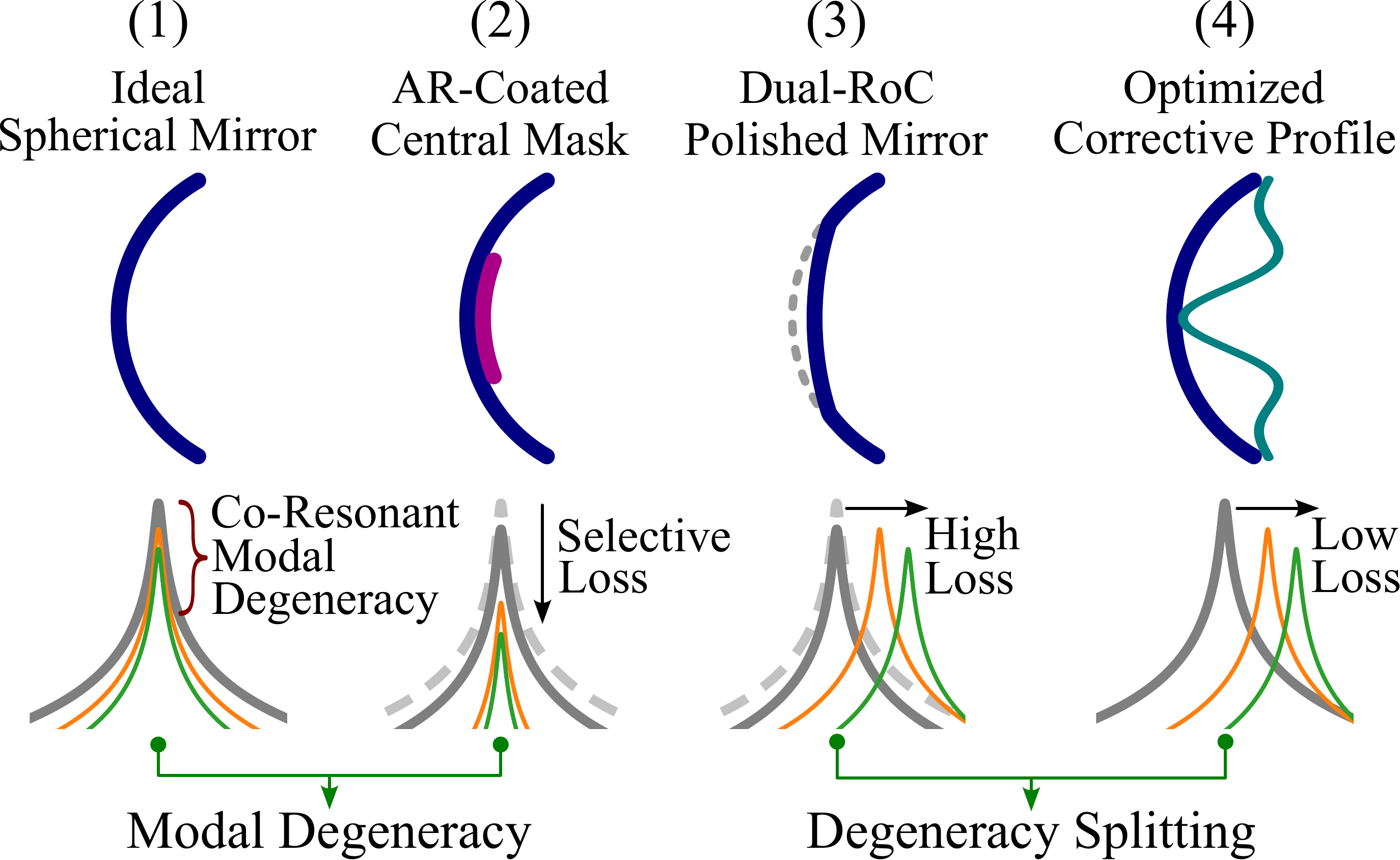}
\caption{Conceptual illustration of modal degeneracy in higher-order LG modes and different approaches for improving the robustness of an injected donut-shaped LG mode in a Fabry-Perot cavity. (1) With ideal spherical mirrors, the target mode, shown in thick gray, is frequency-degenerate with other modes of the same transverse order, shown schematically in orange and green, so that these modes are simultaneously resonant. (2) A central AR mirror mask introduces selective losses for unwanted degenerate modes, but also adds loss to the injected donut-shaped LG mode because of its finite spatial overlap with the masked region. (3) A dual-radius-of-curvature (dual-RoC) mirror profile partially lifts the degeneracy by assigning different curvatures to different radial regions of the test mass, but its non-smooth profile can distort the eigenmode and introduce significant coupling loss. (4) The optimized corrective profile proposed in this work provides a smooth radial surface correction that lifts the modal degeneracy through mode-dependent effective curvatures while preserving the injected donut-shaped LG mode with minimal additional loss.}
\label{fig-LG_illustration}
\end{figure}

To mitigate this limitation, a strategy was recently proposed to improve the robustness of higher-order Laguerre-Gaussian modes against such parasitic imperfections~\cite{tao2026LG06mask}. In particular, donut-shaped \LGmode{0}{\ell} modes have a central intensity minimum and, for sufficiently large $|\ell|$, an extended region of low amplitude around the beam center, making them attractive candidates for reducing test-mass thermal noise while preserving good beam quality. Building on the central intensity minimum and the weak spatial overlap of the donut-shaped \LGmode{0}{\ell} mode with the mirror center, a tailored circular mirror mask with an anti-reflective (AR) coating at the center of the test mass was investigated as a way to introduce selective losses for the unwanted degenerate modes of the same order as the injected donut-shaped \LGmode{0}{\ell} mode. As illustrated in Fig.~\ref{fig-LG_illustration}, the mirror mask suppresses the resonant build-up of parasitic degenerate modes by increasing their optical losses, thereby improving the beam quality and optical performance of the injected mode.

While this approach was effective in improving the injected LG beam quality by reducing both optical loss and contrast defect, it is intrinsically limited by its small design parameter space, which is essentially restricted to the radius of the central AR mirror mask. It therefore does not fully exploit the available degrees of freedom associated with shaping the mirror surface. In addition, because the central mask is a lossy element, it inevitably introduces some loss to the injected mode itself: the overlap between a donut-shaped \LGmode{0}{\ell} mode and the central mask is small, but not strictly zero. For a Virgo-like Fabry-Perot arm cavity, the reflected power loss was substantially reduced and brought closer to the typical values measured in current detectors, but still remained at the few-percent level~\cite{tao2026LG06mask}. Such residual optical loss degrades power coupling into the arm cavities and directly affects the detector sensitivity. The percent level loss is especially important for interferometers employing squeezed-vacuum injection, where optical loss also reduces the observable quantum-noise suppression~\cite{PhysRevD.93.082004, Toyra_2017, McCuller_2021}.

The circular mask increases the purity of the resonant target mode by introducing larger optical losses for the unwanted degenerate modes, thereby suppressing their resonant build-up. An alternative strategy, and the focus of this work, is to suppress the resonant build-up of unwanted modes by lifting the degeneracy within the same mode-order subspace. If the nominally degenerate modes are shifted away from the resonance of the injected target mode, their cavity gain is reduced, so that imperfection-induced scattering into these unwanted modes contributes much less to the circulating optical loss. This principle is closely related to degeneracy lifting previously considered for higher-order Hermite-Gaussian (HG) modes~\cite{Tao2020, PhysRevD.103.042008}. In that case, an astigmatic cavity introduces different Gouy phases along the two transverse Cartesian directions, thereby assigning distinct round-trip Gouy phases to HG modes within the same nominally degenerate subspace. This splits the resonance frequencies of HG modes that would otherwise be degenerate in an ideal optical cavity. The approach is effective because HG modes remain natural eigenmodes of astigmatic cavities, as their field profiles are separable in the transverse Cartesian coordinates; thus, astigmatism does not introduce additional optical loss to the target mode. Once the degeneracy is lifted, modes of the same order no longer co-resonate, making the target HG mode substantially more robust against realistic optical imperfections, with performance comparable to that of the fundamental Gaussian mode~\cite{Tao2020}.

In this work, we extend the concept of resonance-frequency splitting to the Laguerre-Gaussian mode family. For  Laguerre-Gaussian modes, the relevant symmetry is cylindrical rather than Cartesian. The same degeneracy-lifting principle can therefore be applied in a symmetry-matched form: instead of introducing different curvatures along the two transverse Cartesian directions, as in the astigmatic HG case, one can engineer an effective variation of the mirror curvature along the radial direction. This is motivated by the fact that different LG modes within the same mode-order subspace have distinct radial intensity distributions and therefore sample different radial regions of the test-mass surface, as shown in Fig.~\ref{fig-LG_mode_I_phase}. With a suitably designed corrective profile on the test-mass surface, these modes acquire different round-trip phases upon reflection, leading to distinct resonance conditions. As a result, at the resonance of the injected target mode, the unwanted same-order LG modes are shifted away from resonance, and their cavity build-up is suppressed. This preserves the cylindrical symmetry of the target LG mode while mitigating the optical loss associated with imperfection-induced scattering within the originally degenerate subspace.

A simple realization of this radially varying mirror-curvature concept is a specially shaped mirror with distinct radii of curvature (RoCs) in different radial regions. An example is the dual-RoC configuration illustrated in Fig.~\ref{fig-LG_illustration}, where two curvatures are engineered onto the mirror surface: one in the inner circular region and one in the outer annular region, separated by a radial boundary that must be carefully chosen. However, this configuration remains limited by a restricted design space, with essentially only two adjustable parameters: the boundary radius and the curvature difference between the two regions. Moreover, as in the central AR-mask approach, the non-smooth dual-RoC profile can significantly distort the cavity eigenmode away from the injected LG mode, thereby introducing substantial coupling loss for the injected beam, potentially at the percent level.

We can go beyond this simple scheme by directly optimizing a corrective profile on the test-mass surface that simultaneously lifts the degeneracy between LG modes and preserves the eigenmode quality of the injected \LGmode{0}{\ell} mode. The optimized profile is designed to shift the unwanted nominally degenerate modes away from the resonance of the target mode, while minimizing the additional loss introduced to the target mode by the corrective profile itself. We demonstrate this approach for the \LGmode{0}{6} mode in a Virgo-like Fabry-Perot arm cavity. The cavity model is inspired by the arm length and finesse of Advanced Virgo~\cite{Acernese_2015}, while adopting a symmetric mirror geometry chosen for the \LGmode{0}{6} mode. The optimized corrective profile separates the nearest same-order mode, \LGmode{1}{4}, away from the target resonance by $2.04\%$ of the free spectral range (FSR), which is sufficient to reduce the cavity gain of the unwanted modes below unity at the resonance of the injected mode. At the same time, the optimized profile preserves the target mode with essentially no additional loss, while making the \LGmode{0}{6} configuration significantly less sensitive to astigmatism and realistic mirror-surface errors. To quantify this robustness, we perform Monte Carlo simulations using synthetic mirror maps with a root-mean-square (RMS) deformation of $0.1~\mathrm{nm}$, representing an improved surface-quality scenario, and spatial-frequency characteristics representative of measured surfaces, following the procedure of Ref.~\cite{tao2026LG06mask}. Under these conditions, the optimized profile yields an average contrast defect of $368~\mathrm{ppm}$ and an average optical loss of $979~\mathrm{ppm}$. Compared with the central AR-mask method based on selective loss engineering, the optimized profile reduces the average reflected optical loss from the few-percent level to below $0.1\%$, within a factor of two of that obtained with the fundamental Gaussian mode under the same simulation conditions.

The rest of the paper is organized as follows. In \S\ref{sec-methodology}, we introduce the methodology used to design the corrective profile, including the required frequency splitting and the Fourier-Bessel representation of radial surface profiles. In \S\ref{sec-optimization}, we present the optimization strategy, the objective function, and the resulting optimized corrective profile. In \S\ref{sec-result}, we evaluate the performance of the optimized profile in terms of induced mode splitting, profile-induced modal scattering, robustness to astigmatism, Monte Carlo estimates of contrast defect and optical loss, and sensitivity to relative beam-profile offsets. Finally, in \S\ref{sec-conclusion}, we summarize the main results and discuss their implications and possible future extensions.

\section{Methodology \label{sec-methodology}}

\subsection{Frequency Splitting}

The goal of this work is to design an optimized corrective profile on the test-mass surface that separates the resonance of the injected \LGmode{0}{6} mode from those of the other modes in the same nominally degenerate mode-order subspace. Specifically, the induced frequency splitting should be large enough that, at the resonance of the injected mode, the cavity power gain of each unwanted same-order mode is reduced below unity. In this regime, imperfection-induced scattering into these modes is not resonantly enhanced by the optical cavity, thereby limiting their contribution to the circulating optical loss.

The cavity response to a laser field detuned by $\Delta f$ from resonance is modeled using the standard Lorentzian form
\begin{equation}
G(\Delta f) = \frac{G_0} {1+\left(\Delta f/f_{\mathrm{HWHM}}\right)^2},
\label{eq-cavity_lorentzian_gain}
\end{equation}
where $G_0$ is the on-resonance circulating power gain and $f_{\mathrm{HWHM}}$ is the half-width at half-maximum of the resonance. The linewidth is related to the free spectral range and the cavity finesse by
\begin{equation}
f_{\mathrm{HWHM}} = \frac{\mathrm{FSR}}{2\mathcal{F}},
\label{eq-hwhm_gain}
\end{equation}
where $\mathcal{F}$ is the cavity finesse. For a high-finesse Fabry-Perot cavity in the strongly overcoupled limit, as is approximately the case for the arm cavities of ground-based gravitational-wave detectors, the on-resonance circulating power gain is given by
\begin{equation}
G_0 \simeq \frac{2\mathcal{F}}{\pi}.
\label{eq-on_resonance_gain}
\end{equation}

The frequency detuning required for the cavity gain of an unwanted mode to fall below unity is obtained by setting $G(\Delta f)=1$ in Eq.~\eqref{eq-cavity_lorentzian_gain}. This gives
\begin{equation}
\frac{\Delta f}{\mathrm{FSR}} = \frac{1}{2\mathcal{F}}\sqrt{G_0-1}
\simeq \frac{1}{2\mathcal{F}} \sqrt{\frac{2\mathcal{F}}{\pi}-1}.
\end{equation}

For a Virgo-like Fabry-Perot arm cavity with nominal finesse of approximately $\mathcal{F} \simeq 445$~\cite{VirgoCollaboration:25}, this corresponds to a threshold detuning of
\begin{equation}
\frac{\Delta f}{\mathrm{FSR}} \simeq 1.9\times10^{-2}.
\label{eq-detuning_requirement}
\end{equation}
Therefore, the relative detuning between the injected \LGmode{0}{6} mode and each unwanted same-order mode should exceed approximately $1.9\%$ of the FSR to avoid significant resonant enhancement by the cavity.

\begin{figure}[t]
\centering
\includegraphics[width=1\linewidth]{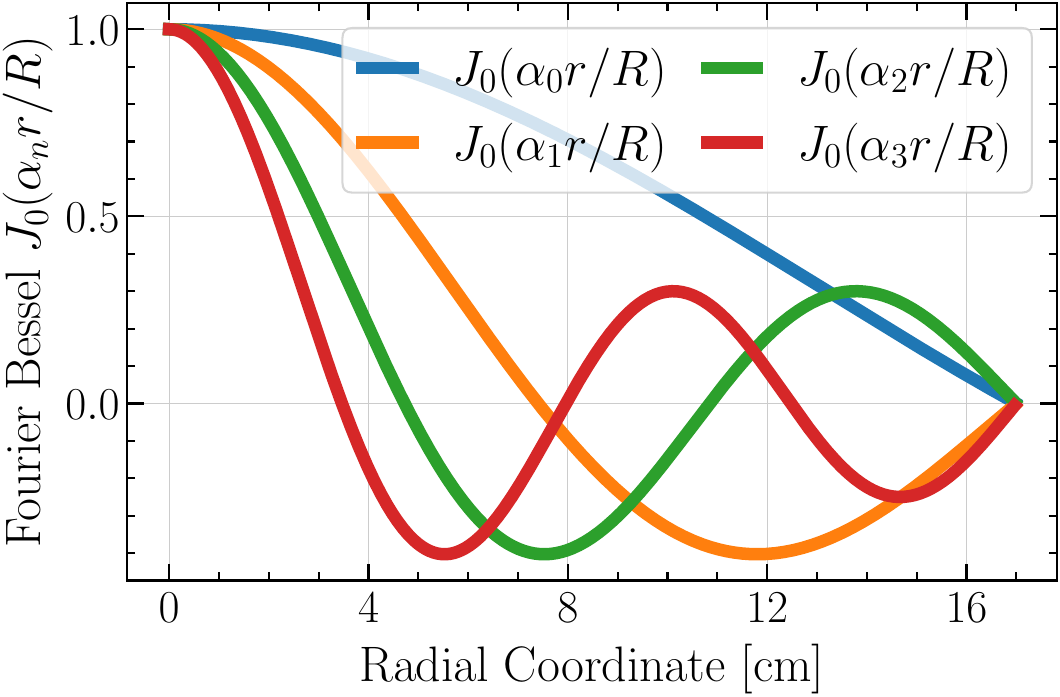}
\caption{First four Fourier-Bessel basis functions used to construct the corrective radial surface-height profile. These smooth basis functions provide sufficient radial structure to tailor the mode-dependent effective curvature sampled by the injected sixth-order \LGmode{0}{6} mode and the other nominally degenerate modes, while avoiding unnecessarily fine radial features.}
\label{fig-fb_basis}
\end{figure}

\subsection{Fourier-Bessel Basis Representation}

The injected \LGmode{0}{6} mode has a circularly symmetric intensity distribution. To lift the degeneracy of the same-order LG modes while preserving the eigenmode quality of the injected mode, we therefore restrict the corrective profile to be circularly symmetric as well. This choice avoids introducing azimuthal asymmetries that would directly scatter power into modes with different azimuthal indices, and reduces the two-dimensional mirror-surface design problem to the optimization of a one-dimensional radial surface profile~\cite{Bond_2011}.

To construct a smooth and physically meaningful radial profile while retaining a sufficiently flexible design parameter space, we expand the corrective surface-height profile in a Fourier-Bessel basis. The Fourier-Bessel functions form a smooth and orthogonal basis for representing circularly symmetric profiles, making them well suited for the present axisymmetric design. We write
\begin{equation}
h(r) = \sum_{n} c_n J_0\left(\alpha_n \frac{r}{R}\right),
\label{eq-fb_profile}
\end{equation}
where $J_0$ is the zeroth-order Bessel function of the first kind, $\alpha_n$ are the zeros of $J_0$, $R$ is the mirror radius, and $c_n$ are the expansion coefficients. For the Virgo-like test mass considered here, we take $R=0.17~\mathrm{m}$. The design of the corrective radial surface-height profile is therefore reduced to the optimization of a finite set of expansion coefficients $c_n$.

The first four Fourier-Bessel basis functions are shown in Fig.~\ref{fig-fb_basis}. Since $J_0(\alpha_n)=0$, each basis function vanishes at $r=R$, providing a natural boundary condition at the mirror edge. Higher-order basis functions contain larger radial spatial-frequency components and therefore enable finer control of the mirror-surface structure. The maximum basis order thus sets the smallest radial features allowed in the corrective profile.

For the donut-shaped \LGmode{0}{6} mode considered in this work, the corresponding sixth-order subspace contains the modes \LGmode{0}{\pm6}, \LGmode{1}{\pm4}, \LGmode{2}{\pm2}, and \LGmode{3}{0}. These modes form four distinct radial intensity profiles, as shown in Fig.~\ref{fig-LG_mode_I_phase}, since each pair \LGmode{p}{+\ell} and \LGmode{p}{-\ell} has the same intensity distribution but opposite phase winding. We find that retaining the first four Fourier-Bessel basis functions provides a suitable balance between design flexibility and optical performance. This expansion allows the corrective profile to address the distinct radial regions sampled by the four intensity distributions in the sixth-order mode subspace, producing differential phase shifts between the distinct radial modes without introducing unnecessarily fine surface features. We assessed this basis truncation by including additional Fourier-Bessel functions and repeating the optimization. Under the settings used here, extending the basis yielded no appreciable improvement in the optimized profile, modal splitting, or target-mode preservation.

\section{Optimization \label{sec-optimization}}

\subsection{Strategy}

The objective of the optimization is to design a corrective surface profile that improves the robustness of an injected donut-shaped LG mode, here taken to be the \LGmode{0}{6} mode. The profile should lift the degeneracy between the target mode and the other modes in the same mode-order subspace, thereby suppressing the resonant build-up of unwanted scattered fields. At the same time, it must preserve the eigenmode quality of the injected mode and avoid introducing additional optical loss from the corrective profile itself.

The optimization therefore balances two competing requirements:
\begin{itemize}
    \item \textbf{Mode splitting:} maximize the resonance frequency separation between the injected \LGmode{0}{6} mode and the unwanted same-order modes, including \LGmode{1}{\pm4}, \LGmode{2}{\pm2}, and \LGmode{3}{0};
    \item \textbf{Mode preservation:} preserve the eigenmode quality of the injected mode and minimize the additional loss induced by the corrective profile.
\end{itemize}

Among the same-order modes, \LGmode{1}{\pm4} has the radial intensity distribution most similar to that of the injected \LGmode{0}{6} mode and therefore has the largest spatial overlap with it. As a result, it samples the corrective profile in the most similar way and typically exhibits the smallest frequency separation from the target mode. Consequently, we use the frequency splitting between the \LGmode{0}{6} and \LGmode{1}{4} modes, denoted by $\Delta f_{(1,4),(0,6)}$, as the representative detuning to be maximized.

Following the weighted-sum method for multi-objective optimization~\cite{marler2010weighted}, this optimization is implemented through the objective loss function
\begin{equation}
\mathcal{L} = -\Delta f_{(1,4),(0,6)} + \eta\,\Gamma_{0,6}^{\mathrm{P}},
\label{eq-objective_function}
\end{equation}
where $\Delta f_{(1,4),(0,6)}$ is the frequency detuning between \LGmode{1}{4} mode and \LGmode{0}{6} mode, expressed in percent of the FSR. The quantity $\Gamma_{0,6}^{\mathrm{P}}$ is the profile-induced reflected power loss of the injected \LGmode{0}{6} mode, expressed in parts per million (ppm), and is defined as
\begin{equation}
\Gamma_{0,6}^{\mathrm{P}} = 1 - \frac{P_{0,6}^{\mathrm{refl.}}}{\widetilde{P}_{0,6}^{\mathrm{refl.}}},
\label{eq-profile_loss}
\end{equation}
where $P_{0,6}^{\mathrm{refl.}}$ denotes the reflected power in the \LGmode{0}{6} mode with the corrective profile, and $\widetilde{P}_{0,6}^{\mathrm{refl.}}$ denotes the corresponding value without the corrective profile. In both cases, the finite mirror aperture with radius $R=0.17~\mathrm{m}$ is included. We consider a symmetric Virgo-like Fabry-Perot arm cavity of length $2998~\mathrm{m}$, with input- and end-test-mass power transmissions of $1.4\%$ and $5~\mathrm{ppm}$, respectively. The nominal mirror radius of curvature is $1596.9~\mathrm{m}$, chosen to set the \LGmode{0}{6} eigenmode beam radius on the test mass to $4.60~\mathrm{cm}$ and to keep the clipping loss at each test mass at $1~\mathrm{ppm}$. The superscript $\mathrm{P}$ indicates that this loss is associated with the corrective profile itself, relative to a reference cavity with the same aperture but without the additional corrective profile.

\begin{figure}[t]
\centering
\includegraphics[width=1\linewidth]{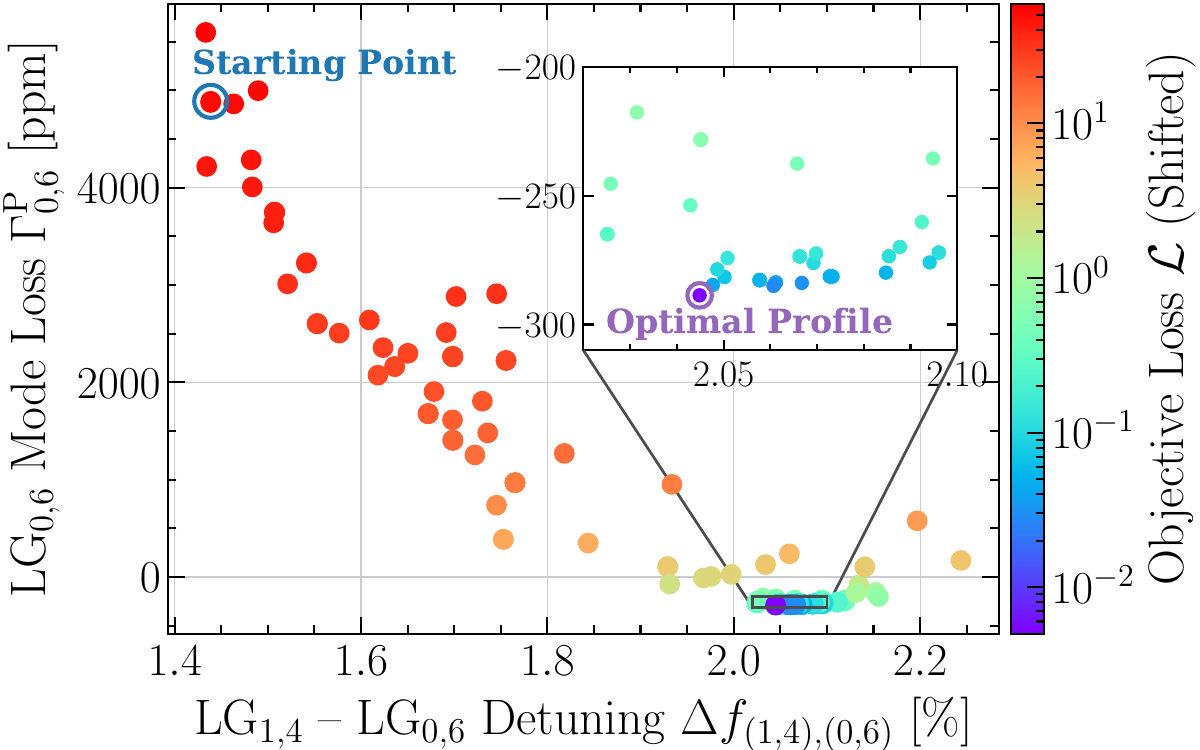}
\caption{Optimization history for the corrective profile design. Each data point represents one optimization step, plotted in terms of the resonance detuning from the nearest same-order mode and the profile-induced loss of the injected \LGmode{0}{6}. The color indicates the value of the objective function. The initial point is labeled ``Starting point,'' and the optimal solution, circled in purple in the inset, achieves large modal detuning while maintaining low profile-induced loss for the injected mode.}
\label{fig-optimization}
\end{figure}

In Eq.~\eqref{eq-objective_function}, the weighting factor $\eta$ controls the trade-off between modal splitting and target-mode preservation. Although the two terms are expressed in different units, the objective function is evaluated using their numerical values in the chosen units. The target detuning is approximately $\Delta f_{(1,4),(0,6)} \simeq 1.9\%$ of the FSR and therefore enters the objective function with a numerical value of approximately $1.9$, i.e., of order unity. By contrast, the optical loss is typically of order several hundred ppm and therefore has a numerical value of order $10^{2}$. We consequently choose $\eta=10^{-2}$ so that the weighted loss term, $\eta\Gamma_{0,6}^{\mathrm{P}}$, is also of order unity and is numerically comparable to the detuning term, $\Delta f_{(1,4),(0,6)}$, during the optimization.

\begin{table}[t]
\centering
\caption{Summary of the optimized profile and key figures of merit. The table lists the optimized Fourier-Bessel coefficients, the modal degeneracy splitting $\Delta f_{(1,4),(0,6)}$, the profile-induced loss of the injected \LGmode{0}{6} mode, and the average contrast defect (CD) and optical loss obtained from randomized Monte Carlo simulations.}
\begin{tabular}{c|c}
\hline
\hline
\textbf{Quantity} & \textbf{Value} \\
\hline
\hline
$c_{n}$ [nm] & $\begin{aligned}[-7.776,\, -11.20,\, -24.34,\, -2.068]\end{aligned}$ \\
\hline
$\Delta f_{(1,4), (0,6)}$/FSR [\%] & 2.04 \\
\hline
Mode Loss $\Gamma_{0,6}^{\mathrm{P}}$ [ppm] & $-288$ \\
\hline
Avg. CD [ppm] & 368 \\
\hline
Avg. Loss [ppm] & 979 \\
\hline
\hline
\end{tabular}
\label{tab-lg06_optimal_profile}
\end{table}

\subsection{Optimal Profile}

The optimization is performed over the first four Fourier-Bessel coefficients $c_n$, with $n=0,1,2,3$. This parametrization provides a sufficiently broad design space while restricting the corrective profile to smooth radial features. The optimization history is shown in Fig.~\ref{fig-optimization}, where each point represents one optimization step and is colored according to the value of the objective function. For visualization on a logarithmic color scale, we plot $\mathcal{L}_{\mathrm{plot}}=\mathcal{L}-\mathcal{L}_{\min}+\epsilon$, where $\mathcal{L}_{\min}=-4.932$ is the minimum objective function value and $\epsilon=0.005$ sets a strictly positive plotting floor. This constant shift preserves the ordering of the objective function values and affects only their visualization, not the optimization. The optimization converges after approximately 200 steps, and the optimal solution, corresponding to the lowest objective value, is indicated by the purple circle in the inset panel.

The optimized Fourier-Bessel coefficients are listed in Tab.~\ref{tab-lg06_optimal_profile}. The resulting resonance frequency splitting between the target \LGmode{0}{6} mode and the nearest same-order mode, \LGmode{1}{4} mode, is $2.04\%$ of the FSR. This exceeds the detuning threshold derived in Eq.~\eqref{eq-detuning_requirement}, ensuring that the cavity gain of the unwanted mode falls below unity at the resonance of the injected mode. In addition, the profile-induced reflected loss of the injected \LGmode{0}{6} mode is $-288~\mathrm{ppm}$. This negative value indicates that, relative to the reference case with the same finite aperture but without the corrective profile, the optimized profile slightly increases the reflected \LGmode{0}{6} power. As discussed in more detail in the next section, this occurs because the profile modifies the effective mirror curvature sampled by the injected mode, leading to a slightly smaller beam size on the mirror and hence reduced clipping loss.

\begin{figure}[t]
\centering
\includegraphics[width=1\linewidth]{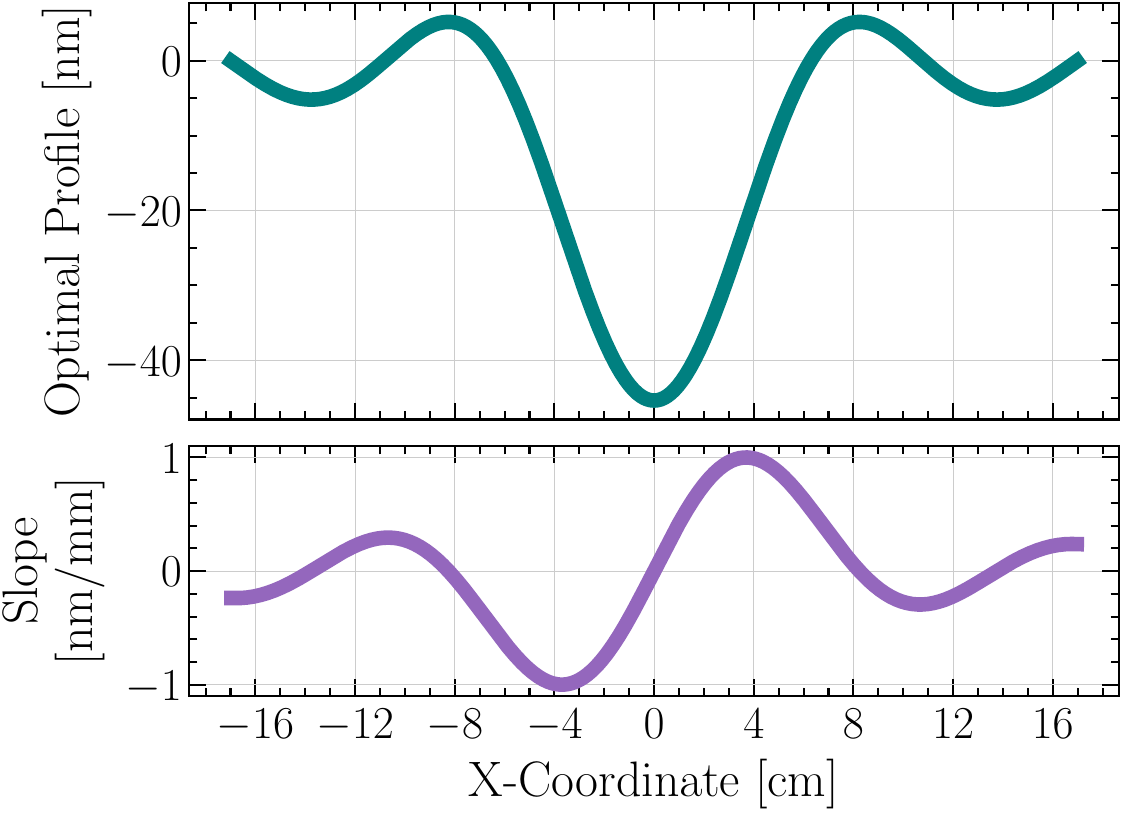}
\caption{Optimized corrective profile. \textit{Top:} Circularly symmetric surface-height profile shown along a horizontal slice through the mirror center. \textit{Bottom:} Radial slope of the corrective profile, expressed as the surface rise per unit radial distance.}
\label{fig-profiles}
\end{figure}

The optimized corrective profile is shown in the upper panel of Fig.~\ref{fig-profiles}. The profile is smooth and radially symmetric, with a peak-to-peak height variation of approximately $50~\mathrm{nm}$. For comparison, the nominal spherical sag of a Virgo-like test mass with radius of curvature $R_{\mathrm{c}}\simeq 1.5~\mathrm{km}$ and radius $R=17~\mathrm{cm}$ is approximately
\begin{equation}
h_{\mathrm{sag}} = R_{\mathrm{c}}-\sqrt{R_{\mathrm{c}}^2-R^2}
\simeq 9.6~\mu\mathrm{m}.
\end{equation}
The optimized corrective profile therefore represents only about $0.5\%$ of the peak-to-peak height variation of the underlying spherical mirror profile.

The optical performance of the corrective surface profile is determined by how its spatial features are sampled by the injected \LGmode{0}{6} mode and by the other modes in the nominally degenerate sixth-order subspace. For instance, a prominent feature of the optimized profile is the radial bump near $r\simeq 8~\mathrm{cm}$, which coincides with the peak-intensity region of the injected \LGmode{0}{6} mode, as shown in Fig.~\ref{fig-LG_mode_I_phase}. The target mode therefore samples this feature strongly, effectively changing the mirror curvature experienced by the injected mode. The other same-order modes have different radial intensity distributions and therefore sample the corrective profile differently. Consequently, the modes across the sixth-order LG subspace experience different effective mirror curvatures and acquire different round-trip phase shifts, leading to distinct resonance conditions and lifting the modal degeneracy.

The lower panel of Fig.~\ref{fig-profiles} shows the corresponding radial slope of the optimized surface profile, expressed as the surface rise in nanometers per millimeter of radial distance. The profile has a peak-to-valley height of approximately $50\,\mathrm{nm}$ and a maximum absolute slope of approximately $1\,\mathrm{nm/mm}$ across the mirror aperture. These modest height and slope variations avoid sharp surface features and are expected to be compatible with standard polishing techniques~\cite{PhysRevD.105.102002, bonnand2012advancedvirgo}.

\section{Results \label{sec-result}}

We now examine the optical implications of the optimized surface profile identified in \S\ref{sec-optimization}. We first evaluate the induced degeneracy lifting across the sixth-order LG subspace and the single-bounce modal scattering introduced by the corrective profile. We then assess the robustness of the optimized profile against astigmatism, realistic randomized mirror surface distortions, and relative beam-profile offsets, using the average contrast defect and optical loss over a large ensemble of trials as the main performance metrics.

\subsection{Mode Degeneracy Lifting}

\begin{figure}[t]
\centering
\includegraphics[width=1\linewidth]{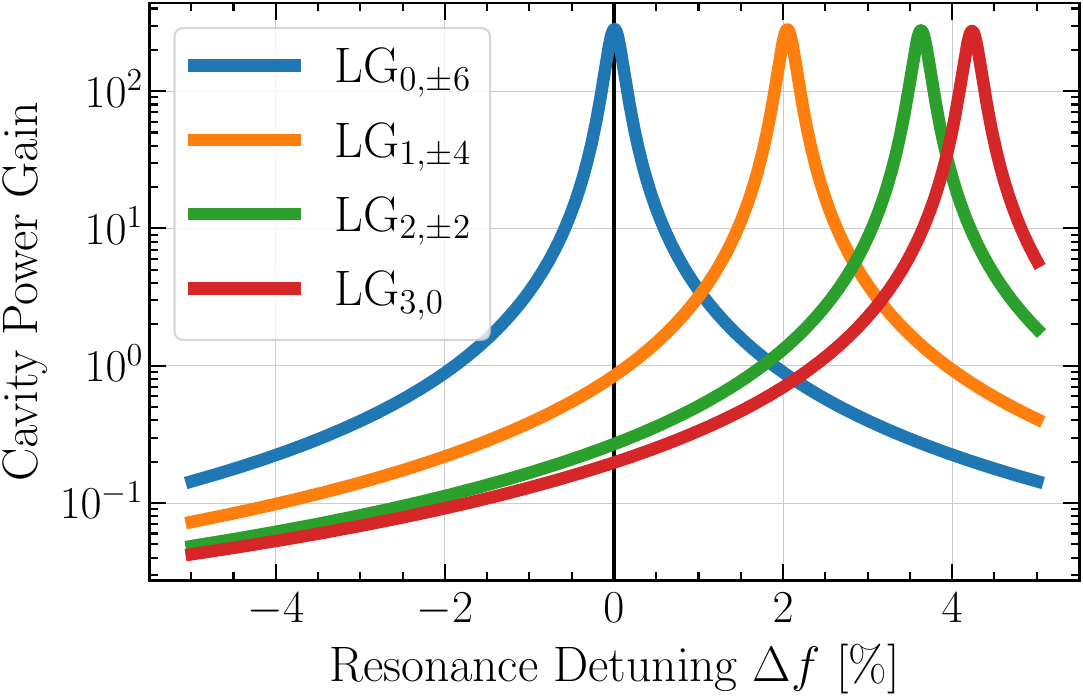}
\caption{Cavity power gain for the same-order $\mathcal{N}=6$ LG modes as a function of frequency detuning relative to the injected \LGmode{0}{6} resonance. The optimized corrective profile lifts the nominal modal degeneracy, shifting the unwanted same-order modes away from the target-mode resonance and reducing their cavity gain at the \LGmode{0}{6} resonance.}
\label{fig-degeneracy}
\end{figure}

Fig.~\ref{fig-degeneracy} shows the cavity power gain scan for the injected \LGmode{0}{6} mode and the other modes in the nominally degenerate sixth-order LG subspace of the Virgo-like Fabry-Perot arm cavity. The results are obtained from optical simulations performed using the interferometer simulation software \textsc{Finesse}~\cite{brown_2025_12662017}. The frequency detuning is shown relative to the resonance of the injected \LGmode{0}{6} mode. Modes of the form \LGmode{p}{+\ell} and \LGmode{p}{-\ell} remain degenerate with one another because they have identical intensity distributions. They therefore sample the circularly symmetric corrective profile in the same way and accumulate the same profile-induced phase shift.

The optimized surface profile produces a clear splitting of the nominally degenerate mode resonances. At the resonance of the injected \LGmode{0}{6} mode, the nearest competing mode, \LGmode{1}{\pm4}, is shifted by more than $2\%$ of the FSR, as listed in Tab.~\ref{tab-lg06_optimal_profile}. Its cavity gain at the injected mode resonance is therefore reduced below unity, satisfying the criterion derived in Eq.~\eqref{eq-detuning_requirement}. This shows that the nearest same-order mode is no longer resonantly enhanced at the target-mode resonance.

The remaining same-order modes, \LGmode{2}{\pm2} and \LGmode{3}{0}, are shifted even further from the injected \LGmode{0}{6} resonance. This larger separation, compared with the \LGmode{1}{\pm4} modes, occurs because their radial intensity distributions differ more strongly from that of the target mode and therefore sample the optimized corrective profile in a more distinct way. Their optical gains at the target-mode resonance are consequently suppressed even further, reaching values below approximately $0.3$. Thus, with the optimized corrective profile, the parasitic same-order modes are no longer resonantly enhanced at the resonance of the injected mode, substantially reducing the scattering loss induced by realistic interferometer imperfections.

\begin{table}[t]
\centering
\caption{Modal decomposition of an injected \LGmode{0}{6} mode after a single reflection from the optimized corrective profile. Only modes with modal power larger than $1~\mathrm{ppm}$ are shown. The modal power is reported in ppm.}
\label{tab-lg_scattering_clean}
\renewcommand{\arraystretch}{1.2}
\setlength{\tabcolsep}{6pt}
\begin{tabular}{l | c}
\hline
\hline
\textbf{Mode} & \textbf{Power [ppm]} \\
\hline

$\mathrm{LG}_{0,6}$ & 998560 \\
$\mathrm{LG}_{1,6}$ & 2 \\
$\mathrm{LG}_{2,6}$ & 878 \\
$\mathrm{LG}_{3,6}$ & 478 \\
$\mathrm{LG}_{4,6}$ & 82 \\

\hline
\multirow{2}{*}{\makecell[c]{\textbf{Total scattered} \\ (non-\LGmode{0}{6})}} 
& 1440 \\
& \textit{(sum over modes)} \\

\hline
\hline
\end{tabular}
\end{table}

\subsection{Profile-Induced Modal Scattering}

To quantify the eigenmode distortion introduced by the optimized corrective profile and the associated profile-induced loss, we compute the modal composition of the field after reflection from the profiled surface. Upon reflection, the corrective surface profile modifies the wavefront of the incident beam. Assuming a normalized cavity eigenmode $u_{p,\ell}(x,y)$, with $(p,\ell)=(0,6)$ in the present study, the reflected field can be written as
\begin{equation}
\widetilde{u}_{p,\ell}(x,y) = u_{p,\ell}(x,y) \exp\left[i\Psi(x,y)\right],
\label{eq-distorted_field}
\end{equation}
where
\begin{equation}
\Psi(x,y) = 2k\,\delta h(x,y).
\label{eq-surface_phase}
\end{equation}
Here, $k=2\pi/\lambda$ is the optical wavenumber, and $\delta h(x,y)$ is the surface-height variation associated with the optimized mirror profile shown in Fig.~\ref{fig-profiles}. The factor of two accounts for the phase shift acquired upon reflection from a surface-height perturbation.

The distorted field $\widetilde{u}_{p,\ell}(x,y)$ can then be expanded in the basis of cavity modes. The complex overlap coefficient with a mode $u_{p',\ell'}(x,y)$ is
\begin{equation}
\begin{aligned}
a_{p',\ell'}
&=
\Braket{u_{p',\ell'} \mid \widetilde{u}_{p,\ell}}  \\
&=
\int
\widetilde{u}_{p,\ell}(x,y)
u_{p',\ell'}^{*}(x,y)
\,\mathrm{d}x\,\mathrm{d}y .
\end{aligned}
\label{eq-mode_overlap}
\end{equation}
The corresponding modal power fraction is therefore
\begin{equation}
P_{p',\ell'}
=
\left|a_{p',\ell'}\right|^2 .
\label{eq-modal_power_fraction}
\end{equation}

\begin{figure}[t]
\centering
\includegraphics[width=1\linewidth]{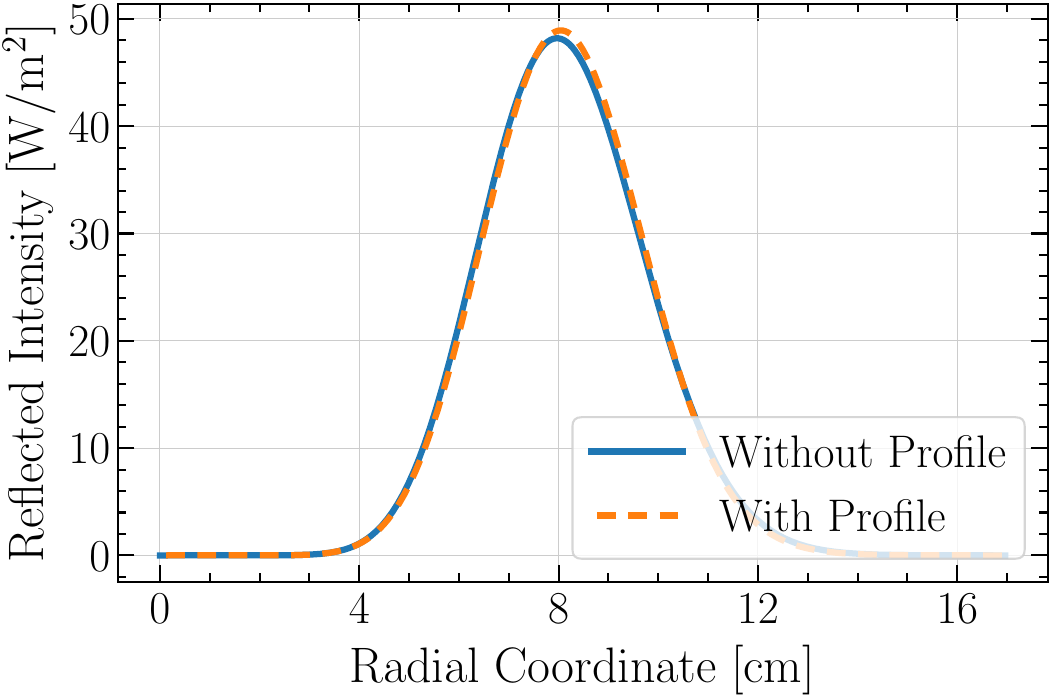}
\caption{Horizontal cross section of the cavity-reflected intensity profile for an injected \LGmode{0}{6} mode, with and without the optimized corrective profile. The corrective profile modifies the effective curvature sampled by the injected mode, in particular through the radial feature near $8~\mathrm{cm}$ where the \LGmode{0}{6} intensity is largest. This leads to a slight reduction of the beam size on the test mass, as indicated by the narrower radial intensity peak and slightly larger peak intensity.}
\label{fig-LG_line}
\end{figure}

The decomposition is performed over scattered modes with arbitrary $p'$ and $\ell'$. Modes with modal power larger than $1~\mathrm{ppm}$ are listed in Tab.~\ref{tab-lg_scattering_clean}, together with their fractional powers. The only modes with significant coupling induced by the optimized profile are \LGmode{p'}{6}-like modes, with $p'=0,1,2,3,4$. This is consistent with the selection rule described by Bond et al.~\cite{Bond_2011}: significant coupling requires $m=|\ell-\ell'|$, where $m$ is the azimuthal index of the relevant Zernike component of the surface distortion, and $\ell$ and $\ell'$ are the azimuthal indices of the injected and coupled modes. Since the optimized corrective profile is axisymmetric, it has $m=0$ and therefore couples the injected \LGmode{0}{6} mode only to modes with the same azimuthal index, namely \LGmode{p'}{6} modes.

After reflection from the corrective profile, the perturbed field remains dominated by the injected \LGmode{0}{6} mode. The total power scattered into all other modes is only $0.14\%$, indicating that the profile preserves high target-mode purity. The dominant scattered components are \LGmode{p'}{6} modes with $p=1,2,3,4$, which have different transverse mode orders from the injected mode. They therefore acquire different Gouy phases and satisfy different resonance conditions, and are not expected to be resonantly enhanced at the \LGmode{0}{6} resonance.

To further examine how the corrective surface profile modifies the reflected field, Fig.~\ref{fig-LG_line} compares the horizontal slices of the intensity at the input port of the Virgo-like arm cavity, with and without the optimized profile applied to both mirrors. In both cases, the cavity-reflected field retains the characteristic single-ring structure of the injected \LGmode{0}{6} mode, with its dominant intensity ring located near $r\simeq 8~\mathrm{cm}$, consistent with Fig.~\ref{fig-LG_mode_I_phase}. Including the corrective profile slightly increases the peak intensity and narrows the radial width of the ring, indicating a slight reduction of the beam size on the test mass.

This change is caused by the radial bump near $r\simeq 8~\mathrm{cm}$ in the optimized profile, shown in Fig.~\ref{fig-profiles}. Because this feature overlaps strongly with the peak-intensity region of the injected \LGmode{0}{6} mode, the target mode samples an effectively smaller curvature of the test mass. Consequently, the modified cavity eigenmode has a slightly smaller beam size on the test mass, reducing the associated optical loss relative to the reference case without the corrective profile. This explains why the profile-induced reflected loss of the target mode can become negative, as reported in Tab.~\ref{tab-lg06_optimal_profile}.

\subsection{Robustness to Astigmatism}

Previous studies have shown that higher-order LG modes are highly sensitive to residual mirror surface distortions, particularly astigmatism at levels compatible with state-of-the-art test-mass polishing~\cite{Sorazu_2013}. This sensitivity arises because astigmatism scatters power from the target mode into other modes within the same nominally degenerate subspace, which are then resonantly enhanced by the cavity. Robustness against such low-order surface perturbations is therefore an essential requirement for any corrective profile designed to improve higher-order LG-mode performance through modal degeneracy lifting.

Because both the injected LG mode and the corrective profile are circularly symmetric, the orientation of the astigmatism can be chosen along the transverse $x$ and $y$ axes without loss of generality. For an astigmatic cavity mirror, we parametrize the curvatures along the two transverse directions as 
\begin{equation} 
\begin{aligned} S_x &= S_0 + \Delta S, \\ S_y &= S_0 - \Delta S, 
\end{aligned} 
\label{eq-astigmatism_curvature} 
\end{equation} 
where $S_x$ and $S_y$ denote the mirror curvatures in the $x$ and $y$ directions, respectively, and $S_0$ is the curvature of the corresponding axisymmetric mirror. The parameter $\Delta S$, expressed in microdiopters ($\mu\mathrm{D}$), therefore quantifies the magnitude of the astigmatic curvature imbalance. For a Virgo test mass with a nominal radius of curvature $R_{\mathrm{c}}\simeq 1500\,\mathrm{m}$, a typical manufacturing astigmatism is approximately $0.01\%$ of the nominal curvature~\cite{bonnand2012advancedvirgo}, corresponding to
\begin{equation}
\Delta S \sim 0.01\% \times \frac{1}{R_{\mathrm{c}}}
\simeq 0.07\,\mu\mathrm{D}.
\end{equation}
Somewhat larger values may arise from coating distortions, suspension-induced deformations, or thermally induced changes in the test-mass surface. In the paraxial approximation, this curvature perturbation corresponds to a surface-height departure $\delta h_{\mathrm{ast}}(r,\phi)=\frac{\Delta S}{2}r^2\cos(2\phi)$ from the nominal axisymmetric surface. Across an aperture of $R_{\mathrm{TM}}=0.17$ m, its peak-to-valley height is
\begin{equation}
h_{\mathrm{PV}}
=|\Delta S|R_{\mathrm{TM}}^2
=28.9\,\mathrm{nm}\,
\frac{|\Delta S|}{1\,\mu\mathrm{D}}.
\label{equ-hPV}
\end{equation}
Thus, for instance, the typical manufacturing astigmatism of $\Delta S=0.07\,\mu\mathrm{D}$ corresponds to approximately 2.0 nm peak-to-valley.

\begin{figure}[t]
\centering
\includegraphics[width=1\linewidth]{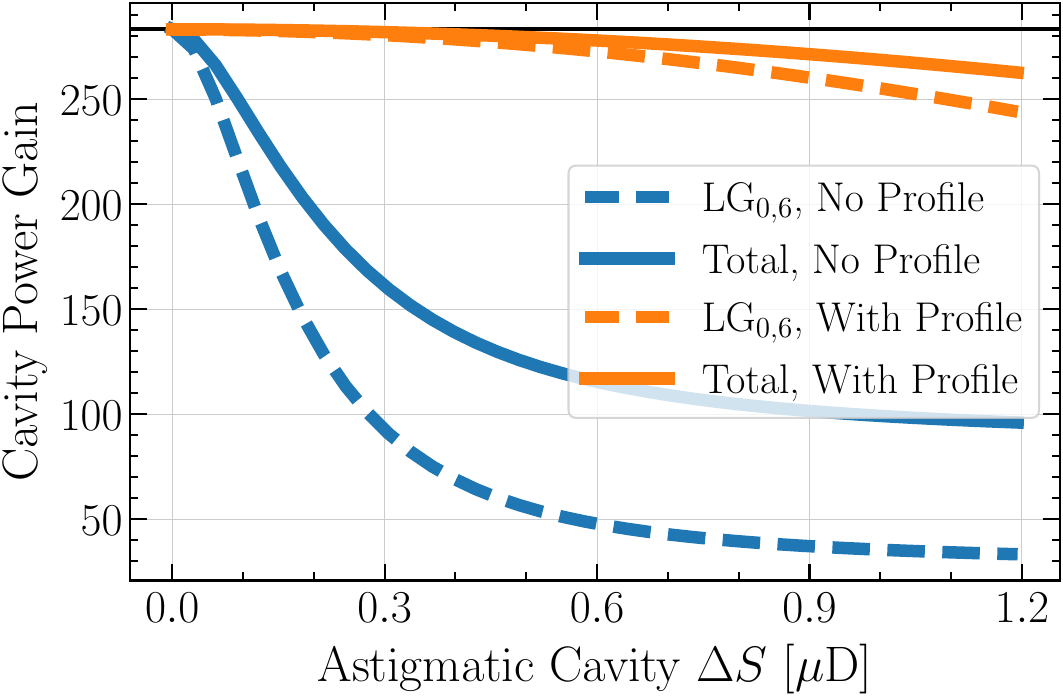}
\caption{Cavity power gain as a function of increasing cavity astigmatism, with and without the optimized corrective profile. The total power gain, including contributions from all modes, is compared with the power gain of the injected \LGmode{0}{6} component alone.}
\label{fig-LG_delta_S}
\end{figure}

Fig.~\ref{fig-LG_delta_S} shows the cavity power gain as a function of the astigmatic curvature imbalance $\Delta S$, applied symmetrically to both mirrors. The gain of the injected \LGmode{0}{6} component is compared with the total power gain summed over all spatial modes present in the system. Both quantities are evaluated at the detuning corresponding to the main cavity resonance peak for each value of $\Delta S$. The optimized profile is applied to both mirrors, together with equal astigmatic curvature perturbations whose principal axes are aligned. The case without the corrective profile is shown in blue, while the case with the optimized profile is shown in orange. Without the corrective profile, both the total power gain and the \LGmode{0}{6} power gain decrease rapidly as the astigmatism increases, illustrating the strong sensitivity of higher-order LG modes to this low-order surface perturbation. In particular, the \LGmode{0}{6} power gain decreases by more than a factor of two, from its nominal value of approximately $283$ to below $140$, already at $\Delta S=0.2~\mu\mathrm{D}$.

By contrast, when the optimized profile is applied, the injected mode gain remains close to its nominal value even in the presence of relatively large astigmatism. At $\Delta S=0.2~\mu\mathrm{D}$, the \LGmode{0}{6} power gain remains above $282$, corresponding to a reduction of only about $0.35\%$. Even for a larger astigmatism of $\Delta S=1~\mu\mathrm{D}$, the mode gain remains above $250$. This demonstrates that the optimized profile substantially suppresses the resonant enhancement of astigmatism-induced scattering and greatly improves the robustness of the injected \LGmode{0}{6} mode.

\begin{figure}[t]
\centering
\includegraphics[width=1\linewidth]{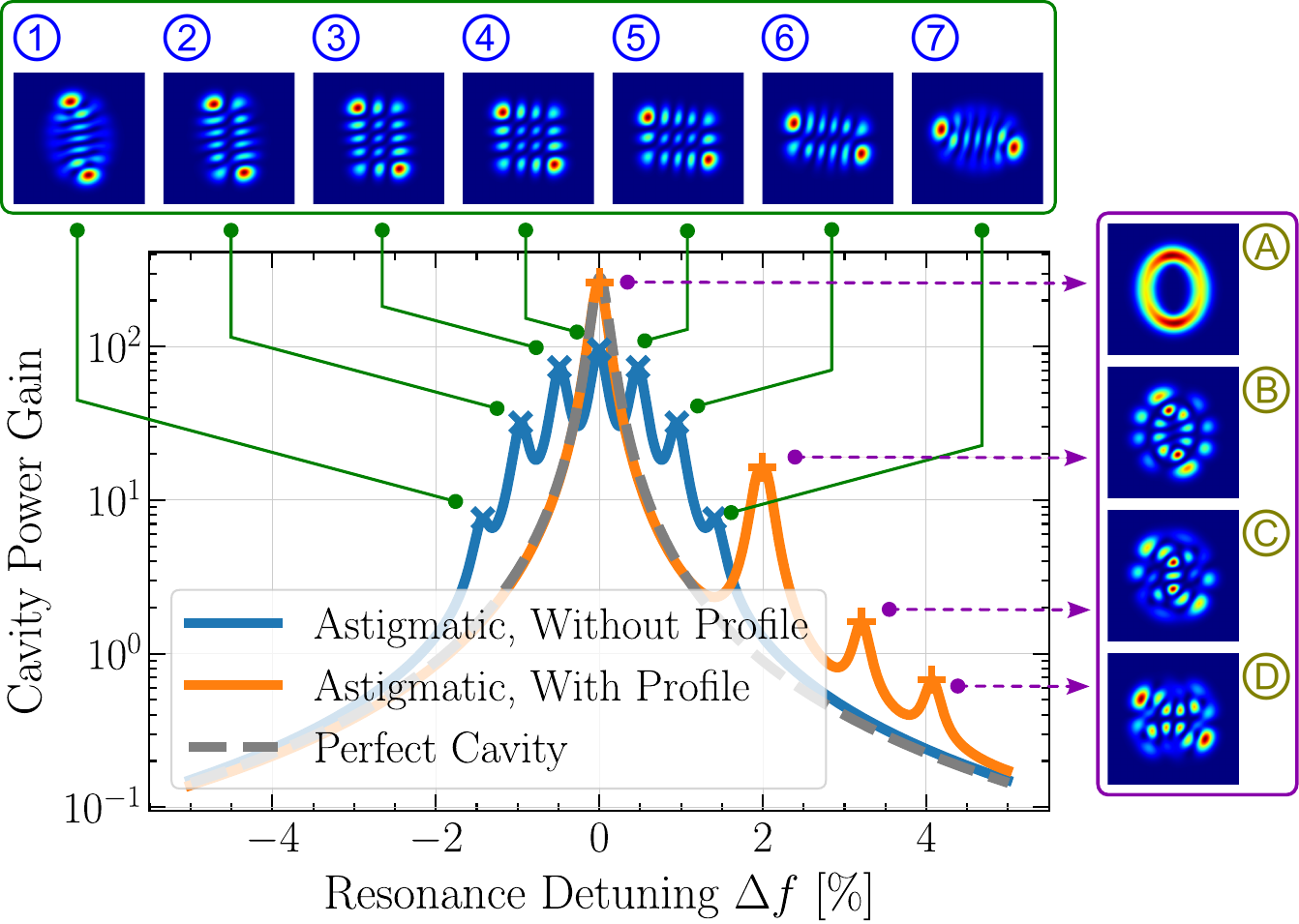}
\caption{Cavity power scan for an injected \LGmode{0}{6} mode with astigmatism $\Delta S = 1.2\,\mu\mathrm{D}$, with and without the corrective profile. The intensity profiles of all visible resonant peaks in the cavity scans are shown.}
\label{fig-LG_astigmatism}
\end{figure}

To further illustrate how the optimized profile preserves the injected mode and reduces the associated loss and impurity in the presence of astigmatism, Fig.~\ref{fig-LG_astigmatism} shows cavity scans for an injected \LGmode{0}{6} mode, with and without the corrective profile. In this example, an astigmatic curvature imbalance of $\Delta S=1.2~\mu\mathrm{D}$ is applied to both cavity mirrors, corresponding to a peak-to-valley surface-height variation of approximately $34.7\,\mathrm{nm}$ across each mirror aperture, as given by Eq.~\eqref{equ-hPV}. Without the corrective profile, the injected \LGmode{0}{6} field is split into seven distinct resonances, corresponding primarily to the same-order Hermite-Gaussian mode basis~\cite{Sorazu_2013}. This is illustrated by the intensity profiles of the seven resonant peaks labeled 1--7. This illustrates the strong modal mixing induced by astigmatism when the order-six mode subspace remains nearly degenerate.

With the optimized profile, the degeneracy between same-order modes is lifted, causing the astigmatism-induced modal components to appear at distinct resonance detunings relative to the target mode. The main resonance associated with the injected \LGmode{0}{6} mode is preserved and retains a high power gain close to the nominal value. The intensity distribution of this main resonance peak, labeled A, remains dominated by the injected \LGmode{0}{6} spatial structure, indicating high modal purity. The additional peaks, labeled B--D, correspond primarily to unwanted same-order resonances that have been shifted away from the target-mode resonance by the optimized profile. These additional resonances exhibit intensity characteristic of sixth-order Ince-Gaussian-like modes, consistent with the eigenmodes of an astigmatic or elliptically distorted cavity~\cite{Schwarz:04}.

\subsection{Monte Carlo Performance}

To evaluate the performance of the optimized corrective profile in the presence of realistic mirror surface errors, we perform Monte Carlo simulations using randomized surface distortion maps with an assumed RMS deformation of $0.1\,\mathrm{nm}$. This value represents an assumed near-future improvement in mirror surface quality that could be enabled by advances in polishing and metrology~\cite{Pinard:17, Billingsley2017_LIGO_P1700029}. The random maps are generated by assigning randomized phases to the two-dimensional spatial-frequency components of a measured reference map, thereby preserving its spatial-frequency characteristics, following the procedure described in Ref.~\cite{tao2026LG06mask}. The measured reference surface used to generate the randomized maps is the LIGO ITM04 mirror surface map. The statistical analysis is performed over many randomized realizations of realistic mirror surface distortions applied to a Virgo-like Fabry-Perot Michelson interferometer, with the two arm cavities modeled using the Virgo-like symmetric Fabry-Perot cavity parameters defined previously. We find that 600 randomized realizations are sufficient for the averaged figures of merit to converge. The average contrast defect and reflected optical loss are used as the main performance metrics, following the definitions in Ref.~\cite{tao2026LG06mask}.

The resulting average contrast defect and optical loss are summarized in Tab.~\ref{tab-lg06_optimal_profile}. For random maps with an RMS deformation of $0.1~\mathrm{nm}$, the optimized corrective profile achieves an average contrast defect of $368~\mathrm{ppm}$ and an average reflected optical loss of $979~\mathrm{ppm}$. The reflected loss is therefore reduced to the $10^{-3}$ level, approaching the performance of the fundamental \LGmode{0}{0} mode, which gives an average loss of approximately $500~\mathrm{ppm}$ under the same conditions~\cite{tao2026LG06mask}. This represents a significant improvement over the central circular AR-mask approach previously proposed in Ref.~\cite{tao2026LG06mask}, for which the reflected loss remains at $4.17\%$, or $2.05\%$ even when the arm cavity finesse is reduced by a factor of two. Reducing the loss from the few-percent level to the $10^{-3}$ level is a key advantage of the optimized corrective profile, particularly for applications with stringent optical loss budgets, such as interferometers using squeezed-light injection.

\begin{figure}[t] 
\centering 
\includegraphics[width=1\linewidth]{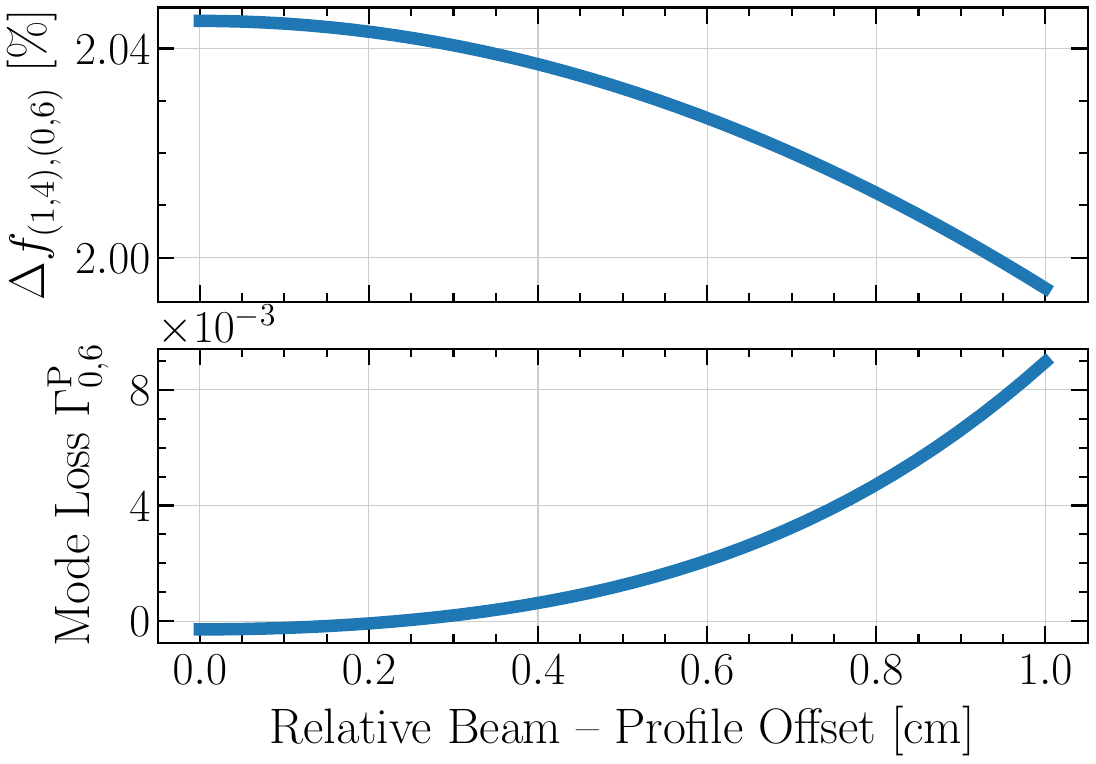} 
\caption{Resonance-frequency splitting and the profile-induced optical loss as a function of the relative offset between the injected mode and the corrective profile.} 
\label{fig-offset} 
\end{figure}

The contrast defect remains larger than that obtained for the fundamental \LGmode{0}{0} mode, for which the average contrast defect is only a few ppm under the same simulated conditions~\cite{tao2026LG06mask}. Nevertheless, the value obtained for the \LGmode{0}{6} mode with the optimized surface profile is comparable to the contrast defect currently measured in Virgo, around $286\pm24~\mathrm{ppm}$~\cite{VirgoCollaboration:25}. It also substantially outperforms the central AR-mask approach, for which the average contrast defect was reported to be $847~\mathrm{ppm}$ under the same simulation assumptions, and $352~\mathrm{ppm}$ after further reducing the arm cavity finesse by a factor of two~\cite{tao2026LG06mask}. Thus, the optimized profile reaches a contrast defect level comparable to the reduced-finesse AR-mask case, but without requiring changes to the arm cavity finesse and the associated retuning of the interferometer response through adjustments to the power- and signal-recycling cavity mirror transmissions.

\subsection{Relative Beam-Profile Offset}

So far, the performance of the optimized corrective profile has been evaluated assuming perfect alignment between the surface profile and the injected \LGmode{0}{6} mode. In practice, however, residual transverse offsets between the optical mode and the corrective profile are unavoidable due to alignment errors, beam jitter, and thermal drifts. It is therefore important to quantify the tolerance of the optimized profile to such relative beam-profile offsets. This sensitivity is a key practical design requirement, since the profile must provide stable performance without imposing overly stringent alignment constraints.

Fig.~\ref{fig-offset} shows the frequency splitting between the injected \LGmode{0}{6} mode and the nearest same-order mode, \LGmode{1}{4}, $\Delta f_{(1,4),(0,6)}$, together with the profile-induced modal loss $\Gamma_{0,6}^{\mathrm{P}}$, as a function of the relative transverse offset between the injected beam and the corrective profile on the end mirror. In this configuration, the beam on the input mirror is assumed to remain perfectly aligned with the corresponding profile, while the profile on the end mirror is laterally displaced relative to the optical mode. As the relative offset increases, the frequency splitting decreases, while the profile-induced loss of the injected mode increases because the displaced profile no longer matches the radial structure of the target mode as effectively. Even for a relative offset as large as $1~\mathrm{cm}$, which is larger than typical alignment offsets expected in practice, the frequency splitting remains above the $1.9\%$ threshold derived in Eq.~\eqref{eq-detuning_requirement}. The resonant gain of the unwanted same-order modes therefore remains strongly suppressed, although the modal loss increases appreciably at large offsets.

\begin{figure}[t] 
\centering 
\includegraphics[width=1\linewidth]{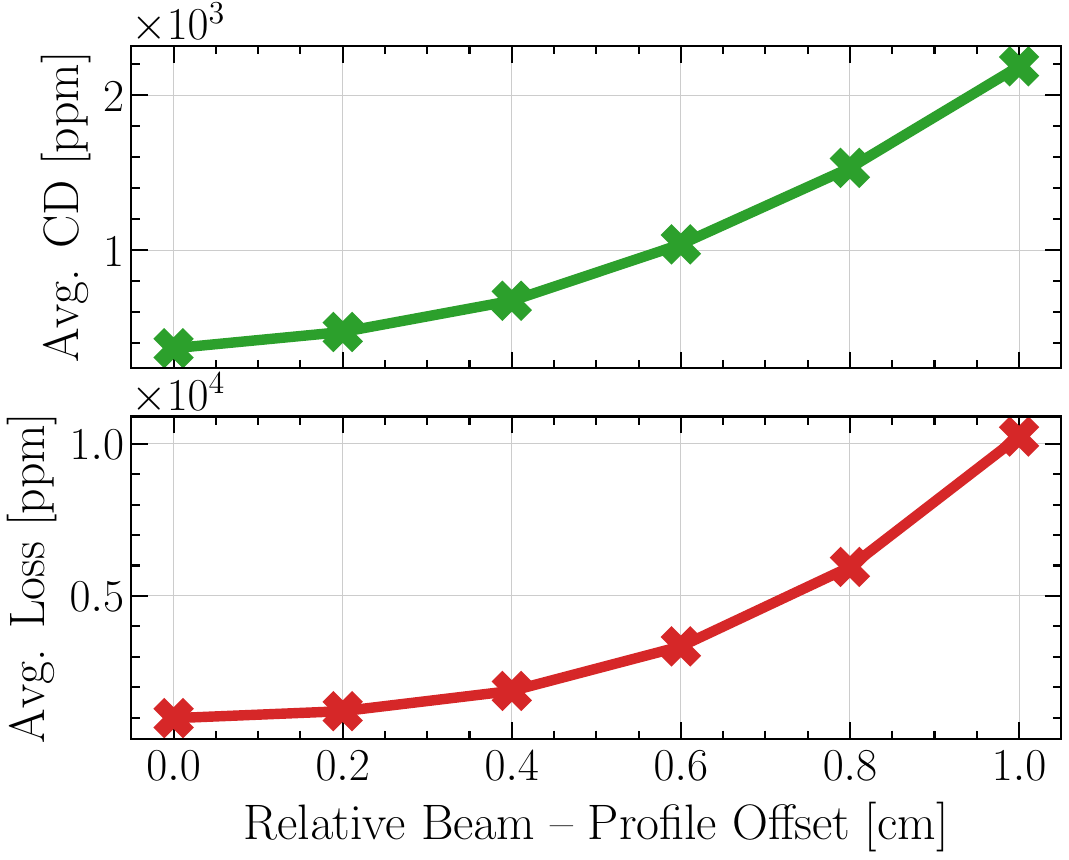} 
\caption{Average contrast defect and reflected optical loss as a function of the relative transverse offset between the injected mode and the optimized corrective profile.} 
\label{fig-offset_avg} 
\end{figure}

Fig.~\ref{fig-offset_avg} shows the corresponding Monte Carlo performance in terms of the average contrast defect and reflected optical loss. To obtain a conservative estimate of the contrast defect, we assume that the relative beam-profile offsets in the two orthogonal arms occur in opposite transverse directions, producing a worst-case mismatch between the reflected fields from the two arms. Both the average contrast defect and the average reflected loss increase with increasing offset. At an offset of $1~\mathrm{cm}$, the average reflected loss reaches the percent level, reducing the advantage of the optimized corrective profile over the central AR-mask approach. Requiring the average contrast defect to remain below $1000~\mathrm{ppm}$ gives a maximum allowable relative offset of approximately $0.6~\mathrm{cm}$. At this offset, the average reflected loss remains around $3000~\mathrm{ppm}$, or $0.3\%$.

The injected \LGmode{0}{6} mode has a beam radius of $w=4.60$ cm on the test masses, with its intensity maximum near $r=8.0$ cm. The largest relative offset considered, 1 cm, is therefore approximately $0.22w$, representing a substantial displacement relative to the beam scale. In practice, mirror alignment and beam spot positions are actively sensed and controlled in precision optical experiments, so relative beam-profile offsets at the level of a few millimeters are expected to be more relevant~\cite{PhysRevD.111.102003}. In this range, the degradation remains small. For example, at a relative offset of $1~\mathrm{mm}$, the performance remains close to the perfectly aligned case, with negligible increases in both the average contrast defect and the reflected optical loss. These results indicate that the optimized corrective profile is tolerant to realistic beam-profile offsets and continues to provide substantial degeneracy lifting and loss reduction under practically relevant alignment conditions.

\section{Discussion and Conclusion \label{sec-conclusion}} 

In this work, we have proposed a general optimization framework for designing corrective surface profiles that modify the mirror curvature to break the modal degeneracy of higher-order Laguerre-Gaussian modes in optical cavities while minimizing the loss of the injected mode. Rather than introducing selective loss through a central AR-coated mirror mask to suppress unwanted same-order modes, the proposed approach directly modifies the cavity eigenstructure by lifting the resonance degeneracy between an injected donut-shaped \LGmode{0}{6} mode and the other unwanted modes in the sixth-order LG subspace. The optimized profile produces a frequency splitting larger than the threshold required to reduce the cavity gain of unwanted modes below unity, while preserving the low-loss character of the injected mode.

The optimized corrective profile is smooth and radially symmetric, with a peak-to-valley height of approximately $50\,\mathrm{nm}$ and a maximum absolute slope of approximately $1\,\mathrm{nm/mm}$, and is expected to be achievable using standard polishing techniques. It produces mode-dependent effective mirror curvatures, allowing same-order LG modes to accumulate different round-trip phases and therefore resonate at different frequencies. At the same time, the profile preserves the modal purity of the injected \LGmode{0}{6} mode, with only minimal modal scattering into higher-radial-order \LGmode{p'}{6} modes. The profile also substantially improves robustness against mirror astigmatism. Monte Carlo simulations with realistic randomized mirror surface distortions show that the optimized profile significantly reduces both the contrast defect and reflected optical loss compared with the central AR-mask approach. In particular, the optical loss is reduced from the few-percent level associated with the central AR-mask approach to the $10^{-3}$ level, approaching the performance of the fundamental Gaussian mode. 

Large-angle scattering arising from high spatial frequency surface features, such as coating nonuniformities and localized point defects, can be estimated by convolving the incident beam intensity distribution with the large-angle scattering loss measured using a scatterometer~\cite{PhysRevD.93.082004, PhysRevApplied.22.054040}. When comparing higher-order LG modes with the fundamental Gaussian mode, only the intensity distribution, which serves as the convolution kernel, changes. Provided that the scattering features are sufficiently localized and broadly distributed across the illuminated region, the resulting large-angle scattering losses are therefore expected to remain of the same order of magnitude. By lifting the modal degeneracy, the optimized corrective profile substantially suppresses small-angle, surface-map-induced modal scattering, reducing the residual average reflected optical loss to $979\,\mathrm{ppm}$. In the small-loss, strongly over-coupled regime, we express this reflected optical loss as an equivalent round-trip loss,
\begin{equation}
L_{\mathrm{rt,eq}} \simeq \frac{\pi}{2\mathcal{F}}\,\Gamma_{\mathrm{refl}},
\label{eq-equivalent_round_trip_loss}
\end{equation}
where $\Gamma_{\mathrm{refl}}$ denotes the fractional reflected optical loss relative to the reference cavity. For $\Gamma_{\mathrm{refl}}=979\times10^{-6}$ and the modeled cavity finesse of approximately $445$, this gives $L_{\mathrm{rt,eq}}\simeq3.5\,\mathrm{ppm}$. This equivalent value characterizes the modeled loss contribution, rather than the complete arm-cavity round-trip loss budget. This value is only approximately twice that of the fundamental mode and remains well below the typical large-angle scattering loss of approximately $20\,\mathrm{ppm}$~\cite{PhysRevApplied.22.054040}. These results suggest that realistic mirror surface imperfections may no longer be a limiting factor in the practical implementation of higher-order \LGmode{0}{\ell}-type modes in laser-interferometric gravitational-wave detectors.

This work represents an important step toward eliminating the excess optical loss associated with higher-order \LGmode{0}{\ell} modes and improving their robustness to realistic optical imperfections, supporting their use for thermal-noise reduction in gravitational-wave detectors. Donut-shaped \LGmode{0}{\ell} modes have also been shown to offer favorable robustness against mode-matching imperfections, further supporting their suitability for practical interferometric implementations~\cite{tao_underpreparation}. The corrective profile studied here is optimized for a specified nominal cavity geometry and injected mode, and the present simulations demonstrate robustness to static astigmatism, randomized surface errors, and relative beam-profile offsets within the investigated ranges.

Although higher-order LG modes are not included in current third-generation detector baseline designs, their potential thermal-noise benefits motivate continued investigation for possible future upgrades. This perspective is particularly relevant to the Einstein Telescope, whose infrastructure is designed to accommodate successive interferometer upgrades over several decades~\cite{ETDesignReportUpdate2020}. The present results contribute to this longer-term assessment by addressing a key practical limitation associated with mirror imperfections. Further detailed simulations are needed to assess the performance of the profiled cavity under general input mode mismatch, cavity-length errors, and time-dependent thermal distortions. Future work will also investigate the full interferometric response, including the impact of the proposed profiles on recycling cavities, optical gain, sensing and control, and quantum-noise reduction in a complete detector configuration. Based on discussions with colleagues at LMA, the amplitude and spatial smoothness of the optimized profile are expected to be compatible with standard polishing techniques. Fabrication and optical characterization of a coated prototype will be needed to verify the achieved surface accuracy and optical performance. A tabletop cavity experiment using such mirrors and a custom-generated donut-shaped LG mode would provide a direct test of both target-mode preservation and the predicted lifting of the modal degeneracy between distinct same-order radial families.

\begin{acknowledgments}
The authors thank Anna Green for helpful comments during the preparation of this manuscript. The authors thank Laurent Pinard and the LMA team for useful discussions. The authors acknowledge the use of the LIGO ITM04 mirror surface map data in generating the randomized mirror maps used in this study. The authors acknowledge support from ANR-23-CE31-0004. This document was submitted to the Virgo and LIGO collaborations under the document numbers VIR-0593A-26 and P2600438, respectively.
\end{acknowledgments}

\bibliography{references}

\end{document}